\documentclass[12pt]{iopart}

\usepackage{iopams}  

\expandafter\let\csname equation*\endcsname\relax

\expandafter\let\csname endequation*\endcsname\relax

\usepackage{amsmath}
\usepackage{cite}
\usepackage{graphicx}
\usepackage{multirow}
\usepackage{bm}

\usepackage{color}

\begin{document}

\title{On dualities for SSEP and ASEP with open boundary conditions}

\author{Jun Ohkubo}
\address{
Department of Computer and Information Science, Saitama University,\\
255 Shimo-Okubo, Sakura-ku, Saitama, 338-8570, Japan
}

\begin{abstract}
Duality relations for simple exclusion processes
with general open boundaries are discussed.
It is shown that a combination of spin operators and bosonic operators
enables us to have an unified discussion for the duality relations
with the open boundaries.
As for the symmetric simple exclusion process (SSEP),
more general results than previous studies are obtained;
it is clarified that not only the absorbing sites,
but also additional sites, called copying sites,
are needed for the boundaries in the dual process for the SSEP.
The role of the copying sites is to conserve information
about the particle states on the boundary sites.
The similar discussions are applied to the
asymmetric simple exclusion process (ASEP),
in which the $q$-analogues are employed,
and it is clarified that the ASEP with open boundaries
has a complicated dual process on the boundaries.
\end{abstract}

\maketitle

\section{Introduction}
\label{sec_introduction}

A concept of duality relations has been widely used
in various research areas.
Especially, in recent years, the duality relations have been used
to investigate various stochastic processes,
ranging from stochastic differential equations
\cite{Shiga1986,Mohle1999,Giardina2007,Carinci2013}
to interacting particle systems,
which include a symmetric simple exclusion process (SSEP)
and an asymmetric simple exclusion process (ASEP)
\cite{Liggett_book,Schutz1997,Imamura2011,Borodin2014,Jansen2014}.
For example, the SSEP and ASEP with reflective boundaries
have self-duality properties,
and it has been shown that
the correlations in the original SSEP and ASEP
are easily investigated
by using the corresponding dual processes
(for example, see \cite{Liggett_book} and \cite{Schutz1997}.)

Although the dual processes and duality functions have been sometimes
derived heuristically, there are a few systematic ways
to investigate the duality relations.
It has been already shown that symmetries of the generators
are useful to derive the duality functions and dual processes
\cite{Giardina2009}.
In \cite{Giardina2009}, the usefulness of the symmetries of the generators
has been demonstrated; the ``classical duality'' (in the sense of \cite{Liggett_book})
has been adequately derived.
However, in general, duality studies with boundary driven cases are difficult.
The SSEP with a specific open boundary condition has been discussed in \cite{Giardina2009},
but the derivation includes some heuristic parts.
Furthermore, as far as I know, the duality relations for the ASEP with open boundaries
has not yet been discussed.
Recently, the ASEP with periodic boundary conditions on a low current
has been discussed \cite{Schutz2015},
but it would be an important remaining task to discuss the open boundary cases. 

In the present paper, a systematic discussion on the duality relations
is given for the SSEP and ASEP with open boundary conditions.
The discussion is based on the symmetries of the quantum Hamiltonian
and the recent developments on the usage of 
bosonic operators, i.e., the so-called Doi-Peliti formalism \cite{Ohkubo2010,Ohkubo2013a}.
Additionally, as for the ASEP,
the $q$-analogues of the exponential functions are employed.

Firstly, the SSEP case is discussed,
and a more general result than the previous works is obtained;
only the absorbing states are needed for the dual stochastic process
in previous studies \cite{Giardina2009},
but additional sites should be used for more general open boundary conditions.
The additional sites are called `copying sites' in the present paper,
and the role is to conserve the particle states on the boundary sites.
Secondly, discussions for the ASEP
with open boundary conditions are given;
from a derivation of the dual time-evolution operator,
it is clarified that the open boundary conditions in the ASEP
give very complicated dual processes. 
This means that the standard duality relations could not be used
for the ASEP case, at least, at this stage.
The current work is the first time to reveal this fact,
by using the systematic discussion proposed in the present paper.

The construction of the present paper is as follows.
In section~\ref{sec_definitions}, 
the formalism, some definitions, and a basic idea to derive duality relations are explained.
Section~\ref{sec_SSEP_1} 
gives a re-derivation of the duality relations
for the SSEP without open boundaries.
The first main contribution of the present paper is given in section~\ref{sec_SSEP_2};
the SSEP with open boundaries are discussed employing the technique 
with bosonic operator formalism,
and general duality relations are derived. 
In section~\ref{sec_ASEP_1}, the previously known duality relations
for the ASEP without open boundaries are re-derived;
this discussion gives us a basis for the cases with open boundaries.
Section~\ref{sec_ASEP_2} is the second main contribution of the present paper;
discussions for the duality relations in the ASEP with open boundaries are given.
Finally, some concluding remarks are denoted in section~\ref{sec_discussions}.

\section{Definitions, notations, and a basic idea}
\label{sec_definitions}

Firstly, some notations based on the quantum Hamiltonian formalism are introduced,
which are useful for the following discussions.
After that, a basic idea to derive the duality relations will be shown.

\subsection{Quantum spin language}

We here employ the following formulation based on the quantum spin language;
for details, see e.g. \cite{Schutz2000}.
We set
\begin{eqnarray}
s^+ = \left( \begin{array}{cc} 0 & 1 \\ 0 & 0 \end{array} \right), \quad
s^- = \left( \begin{array}{cc} 0 & 0 \\ 1 & 0 \end{array} \right), \quad
s^z = \frac{1}{2} \left( \begin{array}{cc} 1 & 0 \\ 0 & -1 \end{array} \right), 
\end{eqnarray}
and the number operator is defined as
\begin{eqnarray}
n = \frac{1}{2}I - s^z =  \left( \begin{array}{cc} 0 & 0 \\ 0 & 1 \end{array} \right).
\end{eqnarray}
Note that the above spin operators obey the following commutation relations:
\begin{eqnarray}
[s^z, s^{\pm}] = \pm s^{\pm}, \quad [s^+, s^-] = 2 s^z.
\end{eqnarray}

\subsection{Time-evolutions for the SSEP and ASEP}

Denote the empty system as $|0\rangle$,
and one can construct an $n$ particle state with particle positions at $x_1, \dots, x_n$ by
\begin{eqnarray}
|x_1, \dots, x_n\rangle = s_{x_1}^- \cdots s_{x_n}^- | 0 \rangle,
\end{eqnarray}
where the operator $s_i^-$ put a particle on site $i$.
On the contrary, the operator $s_i^+$ vanishes a particle on site $i$.
Hence, the operator corresponding to the particle hopping from site $i$ to site $j$ is 
written as $s_j^- s_i^+$.
The state of the system can also be specified by $\eta_i$;
$\eta_i = 1$ (resp. $\eta_i = 0$) means that site $i$ is occupied (resp. empty).
Sometimes the system state is abbreviated as $|\eta \rangle$ with 
$\eta = \{\eta_i | \, i \in \mathcal{S}\}$,
where $\mathcal{S}$ denotes the set of all sites.
Note that the state vector for site $i$ is written
explicitly in terms of vectors;
\begin{eqnarray}
| \eta_i = 1 \rangle = 
\left( \begin{array}{c} 0 \\ 1 \end{array} \right), \quad
| \eta_i = 0 \rangle = 
\left( \begin{array}{c} 1 \\ 0 \end{array} \right).
\end{eqnarray}

In the present paper, we assume that the underlying lattice is a one dimensional one,
that is, a finite lattice $\mathcal{S} = \{ 1,2, \dots, L\}$.
If only the SSEP cases are discussed,
it is possible to deal with arbitrary lattice structures.
However, as we will see in section \ref{sec_ASEP_2},
the discussion for the ASEP relies on the one dimensional structure.
Although it might be possible to extend the following discussions 
for the ASEP on general lattice cases,
it would become very complicated,
and hence the current study focuses only on the one dimensional lattice.

Let $P(\eta,t)$ be the probability that the configuration of the system is $\eta$
at time $t$.
Define
\begin{eqnarray}
|P(t) \rangle = \sum_{\eta} P(\eta,t) | \eta \rangle,
\end{eqnarray}
where $\sum_{\eta}$ means the summation over all particle configurations.
Using a quantum Hamiltonian $H$, 
the time-evolution for $|P(t)\rangle$ in the ASEP is given by
\begin{eqnarray}
\frac{d}{dt} |P(t)\rangle = - H |P(t)\rangle,
\end{eqnarray}
where $H$ is defined as follows: 
\begin{eqnarray}
H = H^\mathrm{bulk} + H^1 + H^L.
\label{eq_H_original}
\end{eqnarray}
Here, $H^\mathrm{bulk}$ denotes the transition matrix
for the bulk part,
and 
\begin{eqnarray}
\fl
H^\mathrm{bulk} =&
- \sum_{k=1}^{L-1} \left[
\alpha_k (s_k^- s_{k+1}^+ - (1-n_k) n_{k+1} )
+ \beta_k (s_k^+ s_{k+1}^- - n_k (1-n_{k+1}))
\right].
\label{eq_H_bulk}
\end{eqnarray}
The open boundary conditions are given by $H^1$ and $H^L$
as follows:
\begin{eqnarray}
H^1 =
- \left[ \gamma^\mathrm{in}_1 s_1^- - \gamma^\mathrm{in}_1 (1-n_1) 
+ \gamma^\mathrm{out}_1 s_1^+ - \gamma^\mathrm{out}_1 n_1
\right], 
\label{eq_H_1} \\
H^L =
- \left[ \gamma^\mathrm{in}_L s_L^- - \gamma^\mathrm{in}_L (1-n_L) 
+ \gamma^\mathrm{out}_L s_L^+ - \gamma^\mathrm{out}_L n_L
\right].
\label{eq_H_L}
\end{eqnarray}
Note that when we set $H = H^\mathrm{bulk}$,
the quantum Hamiltonian gives the ASEP with reflective boundaries.

In the following discussions, according to the previous work in \cite{Schutz1997},
we assume that
\begin{eqnarray}
q_k = \sqrt{\frac{\alpha_k}{\beta_k}} = q, \quad
\mu_k = \sqrt{\alpha_k \beta_k}.
\end{eqnarray}
That is, the asymmetry of the system is uniform,
but the mobility can depend on the lattice sites.

If we set $q = 1$, the Hamiltonian~\eref{eq_H_original}
gives the time-evolution for the SSEP with open boundaries.
Note that the above open boundary conditions
are extensions of the previous work in \cite{Giardina2009}.
In \cite{Giardina2009},
the parameters for the boundaries take a restricted form;
the boundary condition is interpreted as particle reservoirs
whose densities of particles were assumed to be less than one.
Hence, for example, 
the in-rate of particles at boundary site $1$ takes
only $\gamma^\mathrm{in}_1 \in [0,1]$,
and the out-rate was determined as
$\gamma^\mathrm{out}_1 = 1 - \gamma^\mathrm{in}$.
In the present paper, this restriction is not needed.

%
%
%
%

\subsection{Basic idea for the duality relations}

In \cite{Giardina2009}, the duality relations 
have been discussed based on the symmetries of the generators.
In the present paper, a different derivation
using bra-ket notations is employed,
which might be more familiar with many physicists.
Note that the explanation here is only the basic and formal one;
the concrete examples for the derivations
are given in the following sections.

Firstly, we introduce the following dual (bra) vectors:
\begin{eqnarray}
\langle \eta_i' = 1 | =
\left( \begin{array}{cc} 0 & 1 \end{array} \right), \quad
\langle \eta_i' = 0 | = 
\left( \begin{array}{cc} 1 & 0 \end{array} \right),
\end{eqnarray}
where $\langle \eta_i' = 1|$ means that
a particle is on site $i$ in the dual process,
and $\langle \eta_i' = 0|$ corresponds to the absence of a particle
on site $i$.
Hence, for the dual process, 
the spin operator $s_i^+$ makes a particle on site $i$,
and $s_i^-$ annihilates a particle on site $i$;
the role of the spin operators are changed 
compared with the original (ket) vectors.
As in the original process, we abbreviate the particle state of the dual process
as $\langle \eta' |$ with $\eta' = \{ \eta_i' | i \in \mathcal{S} \}$.
In addition, the state vector $\langle P'(t)|$ for the dual process
is defined as
\begin{eqnarray}
\langle P'(t) | = \sum_{\eta'} P'(\eta',t) \langle \eta' |,
\end{eqnarray}
where $P'(\eta',t)$ corresponds to the probability for the dual process
with which the state is $\eta'$ at time $t$.

Note that, at this stage, we have not mentioned about the time evolution
of the dual process; the explicit time-evolution
will be given by using examples in the following discussions.
In addition, if we have the open boundary conditions,
the dual processes should have additional sites;
these extensions are discussed later.

Secondly, choose an time-independent operator $A$, which acts on state vectors.
In general, the operator $A$ does not commute with 
the time-evolution operator $H$ in \eref{eq_H_original}.
Then, the duality relations can be considered as
the explicit expression for the following quantity:
\begin{eqnarray}
\langle P'(t=0) | A | P(t) \rangle &= 
\langle P'(t=0) |  A \rme^{-Ht}  | P(t=0) \rangle \nonumber \\
&= \langle P'(t=0) |  \rme^{-\widetilde{H}t} A  | P(t=0) \rangle \nonumber \\
&= \langle P'(t) | A | P(t=0) \rangle .
\label{eq_basic_idea}
\end{eqnarray}
Here, the new time-evolution operator $\widetilde{H}$ is introduced,
which stems from the interchange of $\rme^{-Ht}$ with $A$.
If $\widetilde{H}$ adequately plays a role
as the time-evolution operator for the dual process,
\eref{eq_basic_idea} becomes the conventional duality relations
between two stochastic processes \cite{Liggett_book};
instead of the time-evolution of the original process,
that of the dual process is available to evaluate
the quantity $A$ in the original process at time $t$.

\section{Re-derivation of duality in the SSEP without open boundaries}
\label{sec_SSEP_1}

We here briefly explain the derivation of the duality relations
in the SSEP without open boundaries, based on the basic idea in \eref{eq_basic_idea}.
In this section, we neglect $H^\mathrm{1}$ and $H^\mathrm{L}$ in \eref{eq_H_original},
and set $H = H^\mathrm{bulk}$.
In addition, we set $q = 1$.

At the beginning, we must choose the operator $A$.
If the operator $A$ is choosen adequately,
we can calculate important quantities for the original SSEP
by solving the dual process.
In addition, for the SSEP problems,
the time-evolution operator has the following special property:
\begin{eqnarray}
H = H^\mathrm{bulk} = (H^\mathrm{bulk})^\mathrm{T}.
\end{eqnarray}
Hence, if we choose the operator $A$ as satisfying the commutative property
with $H = H^\mathrm{bulk}$,
we have $\widetilde{H} = H^\mathrm{bulk}$ in \eref{eq_basic_idea}
and then the dual process obeys the same time-evolutions with the original SSEP.
Because of this self-dual property,
it is expected that the operator $A$ satisfying $[H,A] = 0$
is easy to discuss.

The simplest example is $A = I$, i.e., the identity operator.
In this case, the duality relation reduces to a simple
transition probability;
when $\langle \eta'| = \langle x'_1 \dots x'_n |$
and $| \eta \rangle = | x_1 \dots x_n \rangle$,
this corresponds to the probability
that $n$ particles starting from $x_1, \dots, x_n$
at time 0 are on sites $x'_1, \dots, x'_n$ at time $t$.
In this case, defining the following duality function
\begin{eqnarray}
D(\eta,\eta') = \prod_{i \in \mathcal{S}} \delta_{\eta_i,\eta'_i},
\end{eqnarray}
we have the duality relation
\begin{eqnarray}
\mathbb{E}_\eta \left[ D(\eta_t, \eta') \right]
= \mathbb{E}^{\mathrm{dual}}_{\eta'}
\left[ D(\eta, \eta'_t) \right],
\end{eqnarray}
where 
$\mathbb{E}_{\eta}$ means the expectation for the time-evolution
in the original SSEP process $\eta_t$ starting from the initial state $\eta$;
$\mathbb{E}^\mathrm{dual}_{\eta'}$ corresponds to
the expectation in the dual stochastic process $\eta'_t$
starting from $\eta'$ at $t = 0$.
Of course, it would not be common to call this simplest case as a duality relation.

As for a nontrivial example for the duality relations in the SSEP,
we here choose $A = \exp\left( \sum_{i \in \mathcal{S}} s_i^+ \right)$
(see, for example, \cite{Giardina2009}.)
Note that the operator $A$ is commutative with the time-evolution operator $H$
\cite{Giardina2009};
\begin{eqnarray}
\left[ \exp\left( \sum_{i \in \mathcal{S}} s_i^+ \right) , H \right] = 0.
\end{eqnarray}
In this case, we have the following nontrivial duality function:
\begin{eqnarray}
D(\eta,\eta') = \prod_{i \in \mathcal{S} ; \, \{\eta'_i = 1 \}} \eta_i.
\end{eqnarray}
The product of the above duality function means that 
only if site $i$ in the dual SSEP has a particle,
the product is taken according to the variable $\eta_i$ for the original SSEP.

Based on this nontrivial duality relations,
it is possible to calculate the $m$-th correlation function
for the original SSEP with $n$ particles
by using the dual SSEP with only $m$ particles.
For example, if we want to calculate the 2-body correlation function,
the dual SSEP always has only two particles;
compared with the original SSEP, the dual SSEP is easy to deal with.

Note that the above derivation for the duality relations 
based on the bra and ket notations is essentially 
the same as that based on the symmetries of generators
in \cite{Giardina2009}.
However, we will show that the bra and ket notations is suitable
to extend the discussion for cases with the open boundary conditions.

\section{Duality in the SSEP with open boundaries}
\label{sec_SSEP_2}

\subsection{Non-commutative property of the quantity}

As in the previous section,
the quantity $A = \exp (\sum_{i\in \mathcal{S}} s_i^+ )$
is considered here because this quantity enables us to calculate
the correlation functions.
However, when there are the open boundary conditions,
the operator $A$ does not commute with the time-evolution operator $H$;
\begin{eqnarray}
\left[\exp \left(\sum_{i\in \mathcal{S}} s_i^+ \right), H\right]
&= \left[\exp \left(\sum_{i\in \mathcal{S}} s_i^+ \right), 
H^\mathrm{bulk} + H^\mathrm{1} + H^\mathrm{L}\right]
\nonumber \\
&= \left[\exp \left(\sum_{i\in \mathcal{S}} s_i^+ \right), 
H^\mathrm{1} + H^\mathrm{L}\right] \nonumber \\
&\neq 0.
\end{eqnarray}
Hence, the time-evolution operator $\widetilde{H}$ for the dual process
is different from the original one.

\subsection{Boundary terms and BCH formula}
\label{subsec_BCH}

In order to derive the time-evolution operator $\widetilde{H}$
for the dual stochastic process,
the following Baker-Campbell-Hausdorff (BCH) formula is employed:
\begin{eqnarray}
\rme^{X} \rme^{Y} = 
\exp \left( Y + [X,Y] + \frac{1}{2!}\left[X,[X,Y]\right] + \cdots \right)
\rme^{X}.
\end{eqnarray}
For example, for the boundaries on site $1$, 
\begin{eqnarray}
\left[ s_1^+, H^\mathrm{b} \right]
= - 2 \gamma^\mathrm{in}_1 s_1^z + (\gamma^\mathrm{out}_1 - \gamma^\mathrm{in}_1) s_1^+, \\
\left[ s_1^+, [s_1^+, H^\mathrm{b} ] \right]
= 2 \gamma^\mathrm{in}_1 s_1^+, \\
\left[ s_1^+, [s_1^+, [ s_1^+, H^\mathrm{b} ]] \right]
= 0, 
\end{eqnarray}
and hence after some calculations, we have 
\begin{eqnarray}
\widetilde{H} = H^\mathrm{bulk} + \widetilde{H}^1 + \widetilde{H}^L,
\end{eqnarray}
where
\begin{eqnarray}
\widetilde{H}^1 = -\gamma^\mathrm{in}_1 s^-_1 
+ \left( \gamma^\mathrm{in}_1 + \gamma^\mathrm{out}_1 \right) n_1, 
\label{eq_SSEP_boundary_1}\\
\widetilde{H}^L = -\gamma^\mathrm{in}_L s^-_L
+ \left( \gamma^\mathrm{in}_L + \gamma^\mathrm{out}_L \right) n_L. 
\label{eq_SSEP_boundary_L}
\end{eqnarray}

As discussed above,
the Hamiltonian for the bulk parts has a symmetric property,
i.e., $H^\mathrm{bulk} = (H^\mathrm{bulk})^\mathrm{T}$,
and then the transposed quantum Hamiltonian 
$(H^\mathrm{bulk})^\mathrm{T}$
can be directly interpreted as
the transition matrix for the usual SSEP.
On the other hands,
the boundary parts, $\widetilde{H}^1$ and $\widetilde{H}^L$,
are inadequate as stochastic processes;
there is no probability conservation law.
Hence, it is impossible to consider the operator $\widetilde{H}$
as the time-evolution operator for the dual stochastic process.
In order to recover the characteristics as the stochastic processes,
we need an additional theoretical framework, as described below.

\subsection{Bosonic operators and birth-death processes}

The technique based on the bosonic operators,
the so-called Doi-Peliti formalism \cite{Doi1976,Doi1976a,Peliti1985},
has been mainly used to investigate reaction-diffusion processes
(see, for example, \cite{Tauber2005}.)
The formulation is based on algebraic probability theory
\cite{Hora_Obata_book,Ohkubo2013},
and recently the technique has been used
to derive dual birth-death processes
from stochastic differential equations \cite{Ohkubo2010,Ohkubo2013a}.
It has been shown that 
the bosonic formulation can connect differential operators 
in the Fokker-Planck equations (corresponding to
the stochastic differential equation)
and the creation and annihilation operators in the birth-death processes.

Here, we employ the bosonic operators
in order to deal with the open boundary conditions.
It will be shown that the combination of the spin operators
and the bosonic operators can derive
an adequate dual stochastic process,
even in the case with the open boundaries.

Different from the previous works in \cite{Ohkubo2010,Ohkubo2013a}, 
bra vectors in the Fock space
play important roles to derive the dual process for the SSEP problem.
Firstly, the following bosonic operators and bra vectors
in the Fock space are introduced:
\begin{eqnarray}
\langle \xi' | a^\dagger = \xi' \langle \xi'-1 | , \,\,
\langle \xi' | a = \langle \xi'+1 |,
\end{eqnarray}
where $a^\dagger$ and $a$ are creation and annihilation operators, respectively,
and $\xi' \in \mathbb{N}$.
Note that the roles of `creation' and `annihilation' operators
are changed because we here operate them to the bra vectors;
if we apply them to ket vectors, the names and roles are directly connected.
The bra vector $\langle \xi' |$ corresponds to the number of
particles in a birth-death process.
The vacuum state $\langle 0 |$ is characterized by $\langle 0 | a^\dagger = 0$,
and the creation and annihilation operators satisfy the following commutation 
relations:
\begin{eqnarray}
[a,a^\dagger] \equiv a a^\dagger - a^\dagger a = 1, \,\,
[a,a] = [a^\dagger,a^\dagger] = 0.
\end{eqnarray}
The ket vector $| \zeta \rangle$ is naturally introduced
from the inner product defined as
\begin{eqnarray}
\langle \xi' | \zeta \rangle = \delta_{\zeta,\xi'} \, \xi'!.
\end{eqnarray}

As shown later, in order to deal with the cases with boundary conditions,
the following property of the coherent states is important:
\begin{eqnarray}
a | z \rangle = z | z \rangle,
\end{eqnarray}
where $|z \rangle$ is the coherent state with parameter $z \in \mathbb{R}$,
which is defined as
\begin{eqnarray}
| z \rangle \equiv \rme^{za^\dagger} | 0 \rangle.
\end{eqnarray}

\subsection{Additional sink sites} 

In this subsection, only the case with 
$2-\gamma^\mathrm{in}_i -\gamma^\mathrm{out}_i \neq 0$
($i \in \{1,L\})$
is discussed.
The case with 
$2-\gamma^\mathrm{in}_i -\gamma^\mathrm{out}_i = 0$,
is discussed in Appendix A.

In order to obtain an adequate dual `stochastic' process for the SSEP problem,
we here consider the following state vector $|\widetilde{P}(t)\rangle$:
\begin{eqnarray}
| \widetilde{P}(t) \rangle = \sum_{\eta}
P(\eta,t) | \eta \rangle 
| z^{(1)}_1 \rangle 
| z^{(1)}_L \rangle 
| z^{(2)}_1 \rangle 
| z^{(2)}_L \rangle,
\end{eqnarray}
where $|z^{(1)}_i\rangle$ and $|z^{(2)}_i\rangle$
correspond to the coherent states with parameter $z^{(1)}_i$
and $z^{(2)}_i$, respectively ($i \in \{1,L\}$).
These coherent states, $|z^{(1)}_i\rangle$ and $|z^{(2)}_i\rangle$,
are created by $(a^{(1)}_i)^\dagger$ and $(a^{(2)}_i)^\dagger$,
respectively.
In addition, the corresponding annihilation operators are
$a^{(1)}_i$ and $a^{(2)}_i$.
Corresponding to the `extension' of the state vector $|\widetilde{P}(t)\rangle$,
the following time-evolution operator is introduced:
\begin{eqnarray}
\overline{H} = H^\mathrm{bulk} + \overline{H}^1 + \overline{H}^L,
\label{eq_Hamiltonian_SSEP_DP}
\end{eqnarray}
where
\begin{eqnarray}
\fl
\overline{H}^1 = 
a^{(1)}_1 s_1^- - a^{(1)}_1 (1-n_1) 
+ (2-a^{(1)}_1 - a^{(2)}_1) s_1^+ 
- (2-a^{(1)}_1 - a^{(2)}_1) n_1, \\
\fl
\overline{H}^L = 
a^{(1)}_L s_L^- - a^{(1)}_L (1-n_L) 
+ (2-a^{(1)}_L - a^{(2)}_L) s_L^+ 
- (2-a^{(1)}_L - a^{(2)}_L) n_L, 
\end{eqnarray}
where the annihilation operator $a^{(1)}_i$ (resp. $a^{(2)}_i$) acts only on
the coherent state $| z^{(1)}_i \rangle$ (resp. $|z^{(2)}_i\rangle$).
The reason why we take these forms will become clear below.

If we set $z^{(1)}_i = \gamma^{\mathrm{in}}_i$ and $z^{(2)}_i = 2-\gamma^{\mathrm{in}}_i-\gamma^{\mathrm{out}}_i$
for $i \in \{1,L\}$,
the time-evolution operator \eref{eq_Hamiltonian_SSEP_DP}
becomes the same as the original one in \eref{eq_H_original}
by employing the property of the coherent state.
Here, note that 
$a^{(1)}_i | z^{(1)}_i = \gamma^{\mathrm{in}}_i \rangle = \gamma^{\mathrm{in}}_i | z^{(1)}_i = \gamma^{\mathrm{in}}_i \rangle$
and
\begin{eqnarray}
(2-a^{(1)}_i - a^{(2)}_i) 
| z^{(1)}_i = \gamma^{\mathrm{in}}_i \rangle
| z^{(2)}_i = 2 - \gamma^{\mathrm{in}}_i - \gamma^{\mathrm{out}}_i \rangle \nonumber \\
= (2-\gamma^{\mathrm{in}}_i - (2-\gamma^{\mathrm{in}}_i-\gamma^{\mathrm{out}}_i))
| z^{(1)}_i = \gamma^{\mathrm{in}}_i \rangle
| z^{(2)}_i = 2 - \gamma^{\mathrm{in}}_i - \gamma^{\mathrm{out}}_i \rangle \nonumber \\
= \gamma^{\mathrm{out}}_i
| z^{(1)}_i = \gamma^{\mathrm{in}}_i \rangle
| z^{(2)}_i = 2 - \gamma^{\mathrm{in}}_i - \gamma^{\mathrm{out}}_i \rangle.
\end{eqnarray}

Furthermore, by using the modified quantum Hamiltonian
\eref{eq_Hamiltonian_SSEP_DP},
the basic idea in \eref{eq_basic_idea} gives
the following dual Hamiltonian $\widetilde{H}'$:
\begin{eqnarray}
\widetilde{H}' = H^\mathrm{bulk} + \widetilde{H}'^{1} + \widetilde{H}'^{L}, 
\label{eq_Hamiltonian_SSEP_DP_2}
\end{eqnarray}
where
\begin{eqnarray}
\widetilde{H}'^{1} = 
- \left( a^{(1)}_1 s_1^- - n_1 + a^{(2)}_1 n_1 - n_1 \right), \\
\widetilde{H}'^{L} =
- \left( a^{(1)}_L s_L^- - n_L + a^{(2)}_L n_L - n_L \right).
\end{eqnarray}
(We can easily verify \eref{eq_Hamiltonian_SSEP_DP_2}
by using the correspondences with $\gamma^\mathrm{in}_i \leftrightarrow a^{(1)}_i $
and $2 - \gamma^\mathrm{in}_i - \gamma^\mathrm{out}_i \leftrightarrow a^{(2)}_i$
in \eref{eq_SSEP_boundary_1} and \eref{eq_SSEP_boundary_L}.)
Because the role of the annihilation operators in the dual process
is the creation of the particles,
the above boundary terms are interpreted as follows:
\begin{itemize}
\item[(I)] If a boundary site $i$ has a particle,
the particle can hop into an additional sink site~$1$ attached to 
site $i$ at rate $1$ (this additional sink site stems from the action of $a^{(1)}_i$.)
This is caused by $a^{(1)}_i s_i^- - n_i$.
Hence, we can interpret the sink sites as absorbing ones.
\item[(II)] If a boundary site $i$ has a particle,
the copy of the particle is created, at rate $1$, 
to another additional sink site $2$ attached to site $i$ (which stems from $a^{(2)}_i$.)
Note that, in this case, the particle on site $i$ is not vanished.
This is caused by 
$a^{(2)}_i n_i - n_i$.
In this sence, these sink sites can be interpreted as
`copying' sites.
\end{itemize}

According to the above interpretation of the operators on the boundaries,
the following `extended' state space for the bra vectors
should be used in order to obtain an adequate dual `stochastic' process:
\begin{itemize}
\item Each boundary site $i \in \{1,L\}$ has
two sink sites, i.e., the absorbing and copying ones.
\item The number of particles in the sink sites 
attached to boundary site $i$ are denoted as $\xi'^{(1)}_i \in \mathbb{N}$
(the absorbing sites) and $\xi'^{(2)}_i \in \mathbb{N}$ (the copying sites).
\item Each bra vector $\langle \xi'^{(\l)}_i|$ for $i \in \{1,L\}$
and $\l \in \{1,2\}$ is connected to the creation and annihilation operators
$(a^{(\l)}_i)^\dagger$ an $a^{(\l)}_i$.
\item Initially, the numbers of particles in these sink sites are set to zero;
$\xi'^{(1)}_i = 0$ and $\xi'^{(2)}_i = 0$ for $i \in \{1,L\}$ at $t=0$.
\item The extended dual process has the following state vector,
\begin{eqnarray}
\fl
\langle \widetilde{P}'(t) |
= \sum_{\eta'} \sum_{\xi'^{(1)}_1} \sum_{\xi'^{(2)}_1} 
\sum_{\xi'^{(1)}_L} \sum_{\xi'^{(2)}_L} 
P(\eta',\xi'^{(1)}_1,\xi'^{(2)}_1,\xi'^{(1)}_L,\xi'^{(2)}_L,t) 
\langle \eta'|\langle  \xi'^{(1)}_1| \langle \xi'^{(2)}_1 |
\langle  \xi'^{(1)}_L| \langle \xi'^{(2)}_L |, \nonumber \\
\end{eqnarray}
where $P(\eta',\xi'^{(1)}_1,\xi'^{(2)}_1,\xi'^{(1)}_L,\xi'^{(2)}_L,t)$
is the probability distribution
for the extended dual process.
\item The dual stochastic process obeys the same time-evolution with the original SSEP for the bulk part,
and the time-evolution on the boundaries corresponds to the above explanation (I) and (II).
\end{itemize}

From the identity
\begin{eqnarray}
\langle \xi'^{(\l)}_i  | z^{(\l)}_i \rangle = \left( 
z^{(\l)}_i \right)^{\xi'^{(\l)}_i },
\end{eqnarray}
we finally obtain the following duality function
\begin{eqnarray}
\fl
D\left(\eta,(\eta',\xi'^{(1)}_1, \xi'^{(2)}_1,\xi'^{(1)}_L, \xi'^{(2)}_L)\right) = 
&\left( \prod_{i \in \mathcal{S}; \, \eta'_i = 1} \eta_i \right)
\left( \gamma^\mathrm{in}_1 \right)^{\xi'^{(1)}_1} 
\left(2-\gamma^\mathrm{in}_1-\gamma^\mathrm{out}_1 \right)^{\xi'^{(2)}_1} 
\nonumber \\
& \quad \times
\left( \gamma^\mathrm{in}_L \right)^{\xi'^{(1)}_L} 
\left(2-\gamma^\mathrm{in}_L-\gamma^\mathrm{out}_L \right)^{\xi'^{(2)}_L},
\end{eqnarray}
and the duality relation
\begin{eqnarray}
\mathbb{E}_{\eta}
\left[ 
D\left(\eta_t,(\eta',\xi'^{(1)}_1, \xi'^{(2)}_1,\xi'^{(1)}_L, \xi'^{(2)}_L)\right) 
\right] \nonumber \\
=
\mathbb{E}^\mathrm{dual}_{(\eta',\xi'^{(1)}_1, \xi'^{(2)}_1,\xi'^{(1)}_L, \xi'^{(2)}_L)}
\left[ 
D\left(\eta,(\eta'_t,\xi'^{(1)}_{1,t}, \xi'^{(2)}_{1,t},
\xi'^{(1)}_{L,t}, \xi'^{(2)}_{L,t})\right) 
\right].
\end{eqnarray}
This duality relation is an extension of the previous result in \cite{Giardina2009}.
Note that the above discussions
can adequately recover the previous result in \cite{Giardina2009}
for cases with $\gamma^\mathrm{in}_i = \rho_i$ and 
$\gamma^\mathrm{out}_i = 1 - \rho_i$ ($\rho_i \in [0,1])$;
in this case, only the sink site $1$ for each boundary site $i$
plays the special roles.
That is, the particle copy process does not affect the duality function
because $2-\gamma^\mathrm{in}_i-\gamma^\mathrm{out}_i = 1$.

\section{Re-derivation of the duality in the ASEP without open boundaries}
\label{sec_ASEP_1}

As for the ASEP without open boundaries,
i.e., with reflective boundaries,
the self-dual property has been already derived \cite{Schutz1997}.
In this section, a slightly different derivation is given,
which becomes the basis for the discussions
for the open boundary cases.

\subsection{Similarity transformation and some notations}

Different from the SSEP case,
even in the absence of the open boundary conditions,
the time evolution operator for the dual stochastic process
cannot be obtained easily.
If we choose 
a certain quantity $A$, which is commutative with $H^\mathrm{bulk}$,
we have
\begin{eqnarray}
\langle P'(t=0) | A | P(t) \rangle 
&= \langle P'(t=0) | A \rme^{-H^\mathrm{bulk} t} | P(t=0) \rangle 
\nonumber \\
&= \langle P'(t=0) | \rme^{-H^\mathrm{bulk} t} A | P(t=0) \rangle, 
\label{eq_duality_interchange}
\end{eqnarray}
because $H^\mathrm{bulk} A = A H^\mathrm{bulk}$.
However, 
$H^\mathrm{bulk}$ does not correspond to the time-evolution operator
for the dual stochastic process;
the left action of $-H^\mathrm{bulk}$
does not satisfy the probability conservation law.

In order to recover the probability conservation law,
the following similarity transformation is employed \cite{Schutz1997,Schutz2015}.
Defining 
\begin{eqnarray}
V = q^{\sum_{k=1}^L k n_k},
\end{eqnarray}
it has been shown that the following relation is satisfied:
\begin{eqnarray}
(H^\mathrm{bulk})^\mathrm{T} = V^{2} H^\mathrm{bulk} V^{-2}.
\end{eqnarray}
Note that $\displaystyle \langle P'(t=0) | \rme^{- (H^\mathrm{bulk})^\mathrm{T} t}$
gives an adequate time-evolution of the ASEP for the bra state;
the dual stochastic process obeys the same time-evolutions
with the original ASEP.

Setting the initial states for the bra and ket states as
\begin{eqnarray}
\langle P'(t=0) | \equiv \langle x'_1, \dots, x'_m |, \qquad
| P(t=0) \rangle \equiv | y'_1, \dots, y'_N \rangle,
\end{eqnarray}
we have
\begin{eqnarray}
\fl
\langle x'_1, \dots, x'_m | A | P(t) \rangle 
&= \langle x'_1, \dots, x'_m | V^{-2} 
\rme^{-(H^\mathrm{bulk})^\mathrm{T} t} V^{2} A | P(t=0) \rangle
\nonumber \\
&= q^{-2 \sum_{i=1}^m x'_i} \langle x'_1, \dots, x'_m |
\rme^{-(H^\mathrm{bulk})^\mathrm{T} t} V^{2} A | y'_1, \dots, y'_N \rangle
\nonumber \\
&= q^{-2 \sum_{i=1}^m x'_i} \langle P'(t) |
V^{2} A | y'_1, \dots, y'_N \rangle.
\label{eq_duality_basic}
\end{eqnarray}
As already denoted, 
derived duality function can be varied
depending on the choice of the operator $A$.

Here, for the later use,
let us define the following quantities \cite{Schutz1997}:
\begin{eqnarray}
\fl
S^+ = \sum_{k=1}^L s_k^+(q), \quad
S^- = \sum_{k=1}^L s_k^-(q), \quad
S^z = \sum_{k=1}^L s_k^z = \sum_{k=1}^L \left(\frac{1}{2} I -n_k  \right),
\end{eqnarray}
where
\begin{eqnarray}
\fl
s_k^+(q) = q^{\sum_{j=1}^{k-1} n_j} s_k^+ q^{-\sum_{j=k+1}^L n_j}, \quad
s_k^-(q) = q^{\sum_{j=1}^{k-1} (n_j-1)} s_k^- q^{-\sum_{j=k+1}^L (n_j-1)}.
\end{eqnarray}
In addition, note that the following useful identities: 
\begin{eqnarray}
q^{n_k} s_k^+ = s_k^+, \quad
s_k^+ q^{n_k} = q s_k^+, \quad
q^{n_k} s_k^- = q s_k^-, \quad
s_k^- q^{n_k} = s_k^-.
\end{eqnarray}

\subsection{An example of the duality function}

As for the operator $A$, we here choose $\rme^{S^+}$;
it has been already shown that this quantity can commute
with the quantum Hamiltonian $H^\mathrm{bulk}$ \cite{Schutz1997}.

Define $P'_{\bm{x}}(t)$ as the probability for the ASEP in 
configuration $\bm{x} = \{x_1, \dots, x_m \}$ at time $t$,
and $P_{\bm{y}}(t)$ as that in $\bm{y} = \{y_1, \dots, y_N \}$.
In addition, we here assume that $m < N$.
From \eref{eq_duality_basic}, we have
\begin{eqnarray}
\langle x'_1, \dots, x'_m | \rme^{S^+} | P(t) \rangle
= q^{-2 \sum_{i=1}^m x'_i} \langle P'(t) |
V^{2} \rme^{S^+} | y'_1, \dots, y'_N \rangle.
\label{eq_wo_open_duality_initial}
\end{eqnarray}
Note that the quantum Hamiltonian $H^\mathrm{bulk}$ in \eref{eq_H_bulk}
does not change the total number of particles.
Hence, the total number of particles 
for the ket state $|P(t) \rangle$ is still $N$ even at time $t$
(And of course that for the bra state $\langle P'(t)|$ is $m$.)
Therefore, \eref{eq_wo_open_duality_initial} can be rewritten as
\begin{eqnarray}
\fl
\frac{1}{(N-m)!} 
\sum_{1 \le y_1 < \dots < y_N \le L }
P_{\bm{y}}(t)
\langle x'_1, \dots, x'_m | (S^+)^{N-m}
| y_1, \dots, y_N \rangle \nonumber \\
\fl
=
\frac{1}{(N-m)!} 
q^{-2\sum_{i=1}^m x'_i}
\sum_{1 \le x_1 < \dots < x_m \le L }
P'_{\bm{x}}(t)
q^{2\sum_{i=1}^m x_m}
\langle x_1, \dots, x_m | 
(S^+)^{N-m} | y'_1, \dots, y'_N \rangle.
\label{eq_wo_open_duality_2}
\end{eqnarray}
Here, we used the following facts:
if the bra and ket states have different particle numbers,
the inner product immediately gives zero,
and the operator $S^+$ generates only one particle
to the bra state.
Next, we introduce the projection state,
which has equal weight to any $N$-particle configurations \cite{Schutz1997}:
\begin{eqnarray}
\fl
\sum_{\eta: \sum_{k=1}^L n_k = N} \langle \eta |
\equiv 
\langle N | = \frac{1}{[N]_q!} \langle 0 | (S^+)^N
= \langle 0 | \frac{q - q^{-1}}{q^1 - q^{-1}}
\frac{q - q^{-1}}{q^{2} - q^{-2}}
\dots
\frac{q-q^{-1}}{q^{N} - q^{-N}}
(S^+)^N,
\label{eq_projection_state_1}
\end{eqnarray}
where
\begin{eqnarray}
[x]_q = \frac{q^x - q^{-x}}{q - q^{-1}}
\end{eqnarray}
and
\begin{eqnarray}
[m]_q! = [m]_q \cdot [m-1]_q \cdot \cdots \cdot [1]_q.
\end{eqnarray}
Then, using the following notations, which have been introduced in \cite{Schutz1997},
\begin{eqnarray}
N_x &= \sum_{j=1}^k n_x, \\
Q_x &= q^{-2 N_x}, \\
\widetilde{Q}_x &= \frac{Q_x - Q_{x-1}}{q^{-2}-1} = q^{-2N_{x-1}} n_x
\end{eqnarray}
we have 
\begin{eqnarray}
\langle N | \widetilde{Q}_{x'_1} \cdots \widetilde{Q}_{x'_m}
= q^{-m(N-1)} \langle x'_1, \dots, x'_m |
\frac{(S^+)^{N-m}}{[N-m]_q!}.
\label{eq_bra_Schutz}
\end{eqnarray}
(This identity in \eref{eq_bra_Schutz} has been verified in \cite{Schutz1997}.)
Using these facts, by multiplying an adequate constant,
\eref{eq_wo_open_duality_2}
becomes
\begin{eqnarray}
\fl
 \sum_{1 \le y_1 < \dots < y_N \le L }
P_{\bm{y}}(t)
\langle N | \widetilde{Q}_{x'_1} \cdots \widetilde{Q}_{x'_m}
| y_1, \dots, y_N \rangle \nonumber \\
=
q^{-2\sum_{i=1}^m x'_i}
\sum_{1 \le x_1 < \dots < x_m \le L }
P'_{\bm{x}}(t)
q^{2\sum_{i=1}^m x_m}
\langle N | \widetilde{Q}_{x_1} \cdots \widetilde{Q}_{x_m}
| y'_1, \dots, y'_N \rangle.
\label{eq_wo_open_duality_3}
\end{eqnarray}
\eref{eq_wo_open_duality_3} immediately gives
the duality relation for the ASEP with reflective boundaries,
which has been already derived in \cite{Schutz1997,Schutz2015};
the duality function is given as
\begin{eqnarray}
D(\eta,\eta') = \prod_{i=1}^m q^{- 2 N_{x'_{i}-1} + 2 x'_i} n_{x'_{i}},
\end{eqnarray}
where $n_{x'_i}$ means the number operator for site $x'_i$
in the ket state $\eta$,
and $N_{x'_{i}-1}$ the number of particles
up to site $x'_{i}-1$ in the ket state $\eta$.
(Note that $x'_i$ is the $i$-th particle position in $\eta'$.)

\subsection{Comments for the duality in the reflective boundaries}

In the above discussion,
we selected $\rme^{S^+}$ as the commutative quantity
with $H^\mathrm{bulk}$.
Of course, it is possible to consider different quantities;
similar discussions have been already given
in \cite{Carinci2014}.

Here, there is a comment for the connection with a previous work \cite{Imamura2011}.
In \cite{Imamura2011},
instead of $\rme^{S^+}$,
a slightly different operator,
$\exp(q^{S^z} S^+)$, 
has been used to derive the duality relation for the ASEP with reflective boundaries;
$\exp(q^{S^z} S^+)$ also commutes with $H^\mathrm{bulk}$ \cite{Imamura2011}. 
However, it is possible to show that this different quantity gives
the same duality relation in \eref{eq_wo_open_duality_3};
the derivation is written in Appendix B.
The important point for the derivations is as follows:
for the reflective boundary cases,
the numbers of particles are conserved in the time-evolution
both for the bra and ket states.
In addition, 
the difference between $\rme^{S^+}$ and $\exp\left( q^{S^z} S^+\right)$
is the factor $q^{S^z}$, which depends only on the total number of particles,
and the difference can be removed by multiplying 
certain constants 
for the l.h.s. and r.h.s. in \eref{eq_duality_basic};
as a result, $\rme^{S^+}$ and $\exp\left( q^{S^z} S^+ \right)$
give the same quantity to be calculated.
As we will see later, 
this special characteristic,
i.e., the conservation of the total number of particles,
is not available 
to the open boundary cases.

\section{Duality in the ASEP with open boundaries}
\label{sec_ASEP_2}

In this section, we discuss the duality relation in the ASEP
with open boundaries.
Firstly, some discussions for the commutative quantities with $H^\mathrm{bulk}$
are given. 
Secondly, boundary effects on the dual process are investigated
using $q$-analogues of exponential functions.
The final conclusion is a slightly disappointing one;
the obtained dual process could become very complicated,
and hence, at this stage, the duality relations could not be useful.
In order to find out this fact, the tools developed in the previous sections
are employed; it could be impossible to find the fact
by using a heuristic way.

\subsection{What physical quantities should we use?}

In section~\ref{sec_ASEP_1}.3, 
we discussed that two different types of quantities,
$\rme^{S^+}$ and $\exp( q^{S^z} S^+)$,
give the same duality relation.
Here, we will show that the similar discussion
\textit{cannot} be used for the open boundary cases.

For the open boundary cases,
there is no guarantee that
the number of particles is conserved in the processes;
because of the in- and out-effects on the boundaries (site $1$ and $L$),
the number of particles can vary with time.
Hence, the l.h.s. in \eref{eq_wo_open_duality_initial} becomes 
\begin{eqnarray}
\fl
\langle x'_1, \dots, x'_m | \rme^{S^+} | P(t) \rangle
= \sum_{N=0}^\infty
\sum_{n=0}^\infty 
\sum_{1 \le y_1 < \dots < y_N \le L} 
\langle x'_1, \dots, x'_m |
\frac{1}{n!} (S^+)^n P_{\bm{y}}(t)
| y_1 , \dots, y_N \rangle .
\end{eqnarray}
Note that the summation for $N$ is necessary,
different from section~\ref{sec_ASEP_1}. 

After some calculations, we have
\begin{eqnarray}
\fl
\langle x'_1, \dots, x'_m | \rme^{S^+} | P(t) \rangle \nonumber \\
\fl
=\sum_{N=m}^\infty
\sum_{1 \le y_1 < \dots < y_{N} \le L} 
\frac{q^{m(N-1)}[N-m]_q!}{(N-m)!}
P_{\bm{y}}(t)
\langle N | \widetilde{Q}_{x'_1} \cdots \widetilde{Q}_{x'_m}
| y_1 , \dots, y_{N} \rangle,
\end{eqnarray}
but this does not give the expectation for 
$\widetilde{Q}_{x'_1} \cdots \widetilde{Q}_{x'_m}$;
the coefficients
$\frac{q^{m(N-1)}[N-m]_q!}{(N-m)!}$ remain, 
and then a kind of \textit{weighted} expectation is obtained.

In order to obtain an usual expectation for 
$\widetilde{Q}_{x'_1} \cdots \widetilde{Q}_{x'_m}$,
the following quantity, 
which is based on the $q$-analogue of exponential functions \cite{Koekoek_book},
is available:
\begin{eqnarray}
A = \mathrm{e}_{q^{-2}} \left( (1-q^{-2})q^{-\sum_{k=1}^L n_k} S^+ \right),
\label{eq_def_quantity_A}
\end{eqnarray}
where 
\begin{eqnarray}
\mathrm{e}_q (x) \equiv \sum_{n \ge 0} \frac{x^n}{ (q;q)_n }, \qquad 0 < | q | < 1, 
\quad | x | < 1,
\label{eq_def_q_exponential_1}
\end{eqnarray}
and
\begin{eqnarray}
(q;q)_n \equiv \prod_{i=1}^n (1-q^{i}).
\end{eqnarray}
Note that $q^{S^z} S^+$ commutes with $H^\mathrm{bulk}$,
and then we have
\begin{eqnarray}
\left[ \mathrm{e}_{q^{-2}} \left( (1-q^{-2})q^{-\sum_{k=1}^L n_k} S^+ \right),
H^\mathrm{bulk} \right]
= 0.
\end{eqnarray}
As shown in Appendix C,
this quantity gives the usual expectation for 
$\widetilde{Q}_{x'_1} \cdots \widetilde{Q}_{x'_m}$;
\begin{eqnarray}
\fl
\langle x'_1, \dots, x'_m | \mathrm{e}_{q^{-2}}
\left( (1-q^{-2})q^{-\sum_{k=1}^L n_k} S^+ \right)
| P(t) \rangle \nonumber \\
=\sum_{N=m}^\infty 
\sum_{1 \le y_1 < \dots < y_N \le L} 
\langle N |
\widetilde{Q}_{x'_1} \cdots \widetilde{Q}_{x'_m}
P_{\bm{y}}(t)
| y_1 , \dots, y_N \rangle.
\label{eq_w_open_quantity}
\end{eqnarray}

Notice the following facts:
in order to use the $q$-analogues,
we must restrict the following discussions
for the cases with $q \ge 1$,
because $\lvert q^{-2} \rvert < 1$ is needed
for the definition of the $q$-analogues.
(The $q = 1$ case immediately reduces to the SSEP case, 
and the following discussions can be easily obtained
by using the usual exponential functions.
The following formulations are formally available 
even in the $q=1$ case,
and then we here consider $q \ge 1$.)
That is, $\alpha_k > \beta_k$ for all $k \in \mathcal{S}$.
The discussions for $q < 1$ cases
need the change of the order of the lattice structures.

\subsection{Effects of the open boundaries on the dual process}

In order to discuss the duality relations,
the following different type of $q$-analogues of the exponential functions
is useful \cite{Carinci2014}:
\begin{eqnarray}
\exp_q(x) \equiv \sum_{n\ge 0} \frac{x^n}{\{n\}_q!},
\label{eq_def_q_exponential_2}
\end{eqnarray}
where
\begin{eqnarray}
\{n\}_q \equiv \frac{1-q^n}{1-q},
\end{eqnarray}
and
\begin{eqnarray}
\{n\}_q! \equiv \{n\}_q \cdot \{n-1\}_q \cdot \cdots \cdot \{1\}_q.
\end{eqnarray}
That is, two types of the $q$-exponentials in \eref{eq_def_q_exponential_1}
and \eref{eq_def_q_exponential_2}
are related each other as follows:
\begin{eqnarray}
\mathrm{e}_{q^{-2}} ((1-q^{-2})z)
= \exp_{q^{-2}} (z).
\end{eqnarray}

For the open boundary cases,
we must consider the quantum Hamiltonian including $H^1$ and $H^L$.
As for the quantity $A$ in \eref{eq_def_quantity_A},
we have
\begin{eqnarray*}
\left[\exp_{q^{-2}} \left( (1-q^{-2})q^{-\sum_{k=1}^L n_k} S^+ \right), H \right]
\neq 0.
\end{eqnarray*}
Different from the SSEP cases in section~\ref{sec_SSEP_2},
it is impossible to employ the BCH formula directly,
because we here use the $q$-analogues of the exponential functions.
Hence, if we want to perform the similar discussions
in \eref{eq_duality_interchange},
it is necessary to seek the following alternative quantum Hamiltonian $\widetilde{H}$:
\begin{eqnarray}
\exp_{q^{-2}} \left( q^{-\sum_{k=1}^L n_k} S^+ \right) H
= \widetilde{H} \exp_{q^{-2}} \left( q^{-\sum_{k=1}^L n_k} S^+ \right).
\end{eqnarray}
Then, we have
\begin{eqnarray}
\exp_{q^{-2}} \left( q^{-\sum_{k=1}^L n_k} S^+ \right) \rme^{-Ht}
= \rme^{-\widetilde{H}t}\exp_{q^{-2}} \left( q^{-\sum_{k=1}^L n_k} S^+ \right).
\end{eqnarray}

In the discussions for the effects of the open boundaries,
the following proposition (Proposition 5.1 in \cite{Carinci2014}) is useful:
\vskip 3mm
\textbf{Proposition} (Pseudo-factorization \cite{Carinci2014})
{\it 
Let $\{g_1, \dots, g_L\}$ and $\{k_1, \dots, k_L\}$
be operators such that for $L \in \mathbb{N}$ and $r \in \mathbb{R}$
\begin{eqnarray}
k_i g_i = r g_i k_i \qquad \textrm{for} \quad i = 1, \dots, L.
\end{eqnarray}
Define
\begin{eqnarray}
\hat{g}^{(L)} \equiv \sum_{i=1}^L g_i h^{(i+1)},
\end{eqnarray}
with
\begin{eqnarray}
h^{(i)} \equiv k_i^{-1} \cdots k_L^{-1} \qquad \textrm{for} \quad i \le L
\quad \textrm{and} \quad h^{(L+1)} = 1,
\end{eqnarray}
then
\begin{eqnarray}
\exp_{r}\left( \hat{g}^{(L)} \right)
= \exp_r \left( g_1 h^{(2)} \right)
\cdots \exp_r \left(g_{L-1} h^{(L)}\right) \cdot \exp_r \left(g_L\right).
\end{eqnarray}
}
\vskip 3mm

Now, setting
\begin{eqnarray}
g_i = s^+_i, \quad k_i = q^{2n_i}, \quad r = q^{-2},
\end{eqnarray}
then
\begin{eqnarray}
q^{2n_i} s^+_i = q^{-2} s^+_i q^{2n_i},
\end{eqnarray}
and
\begin{eqnarray}
g_i h^{(i+1)} = g_i k^{-1}_{i+1} \cdots k^{-1}_L
= s^+_i q^{-2n_{i+1}} \cdots q^{-2n_L}
= s^+_i q^{-2\sum_{j=i+1}^L n_j}.
\end{eqnarray}
Therefore, we have the following factorization:
\begin{eqnarray}
\fl
\exp_{q^{-2}} \left( \sum_{k=1}^L s_k^+ q^{-2\sum_{j=k+1}^L n_j} \right) \nonumber \\
= \exp_{q^{-2}} \left(s_1^+ q^{-2\sum_{j=2}^L n_j}\right)
\exp_{q^{-2}} \left( s_2^+ q^{-2\sum_{j=3}^L n_j}\right)
\cdots \exp_{q^{-2}} \left( s_L^+ \right).
\label{eq_factorization}
\end{eqnarray}

Because
\begin{eqnarray}
\left[\exp_{q^{-2}} \left( (1-q^{-2})q^{-\sum_{k=1}^L n_k} S^+ \right), H^\mathrm{bulk} \right]
= 0,
\end{eqnarray}
it is enough to consider the interchange with 
$\exp_{q^{-2}} \left( (1-q^{-2})q^{-\sum_{k=1}^L n_k} S^+ \right)$
and $H^1$ (and $H^L$).

\subsubsection{Discussion for site $1$} 

The aim here is to seek $\widetilde{H}^1$, which is obtained by
\begin{eqnarray}
\exp_{q^{-2}} \left( q^{-\sum_{k=1}^L n_k} S^+ \right) H^1
= \widetilde{H}^1 \exp_{q^{-2}} \left( q^{-\sum_{k=1}^L n_k} S^+ \right).
\end{eqnarray}
In the r.h.s.~in \eref{eq_factorization},
only the first factor,
$\exp_{q^{-2}} \left(s_1^+ q^{-2\sum_{j=2}^L n_j}\right)$,
does not commute with $H^1$.
Hence, we focus on the interchange between 
$\exp_{q^{-2}} \left(s_1^+ q^{-2\sum_{j=2}^L n_j}\right)$
and
$\exp_{q^{-2}} \left( q^{-\sum_{k=1}^L n_k} S^+ \right)$.

For notational simplicity, let us define
\begin{eqnarray}
\zeta = q^{-2 \sum_{j=2}^L n_j}.
\end{eqnarray}
Then, because of $s^+_1 s^+_1 = 0$,
\begin{eqnarray}
\fl
\exp_{q^{-2}} \left( s_1^+ q^{-2 \sum_{j=2}^L n_j} \right) 
= \sum_{n=0}^\infty \frac{1}{\{n\}_{q^{-2}}!} 
\left( s_1^+ q^{-2 \sum_{j=2}^L n_j}  \right)^n
= I +  s_1^+ \zeta.
\end{eqnarray}
In addition, introducing 
\begin{eqnarray}
\mathrm{E}_q (z) \equiv \sum_{n=0}^\infty
\frac{q^{\binom{n}{2}}}{(q;q)_n} z^n,
\end{eqnarray}
where
\begin{eqnarray*}
\binom{\alpha}{\beta}
= \frac{\Gamma(\alpha+1)}{\Gamma(\beta+1)\Gamma(\alpha-\beta+1)},
\end{eqnarray*}
we have \cite{Koekoek_book}
\begin{eqnarray}
\rme_q(z) \mathrm{E}_q(-z) = 1.
\end{eqnarray}
Using this `inverse' function of $\rme_q(z)$,
we can factor out $\exp_{q^{-2}}(s^+_1 \zeta)$.
Hence, after some tedious calculations, the following
factorized form using $\exp_{q^{-2}}(s^+_1 \zeta)$ is obtained:
\begin{eqnarray}
\fl
\exp_{q^{-2}} \left( s_1^+ \zeta \right) H^1 \nonumber \\
\fl
= \left(
-\gamma^\mathrm{in}_1 s_1^- 
+ [\gamma^\mathrm{in}_1 \zeta^2 - \gamma^\mathrm{out}_1 
- (\gamma^\mathrm{in}_1 - \gamma^\mathrm{out}_1)\zeta]s_1^+
+ (\gamma^\mathrm{in}_1 - \gamma^\mathrm{in}_1 \zeta) I
+ [ - (\gamma^\mathrm{in}_1 - \gamma^\mathrm{out}_1) + 2 \gamma^\mathrm{in}_1 \zeta ] n_1
\right) \nonumber \\
\fl \quad \times
\exp_{q^{-2}} \left( s_1^+ \zeta \right) ,
\end{eqnarray}
and then
\begin{eqnarray}
\fl
\widetilde{H}^1 =& 
-\gamma^\mathrm{in}_1 s_1^- 
+ [\gamma^\mathrm{in}_1 \zeta^2 - \gamma^\mathrm{out}_1 
- (\gamma^\mathrm{in}_1 - \gamma^\mathrm{out}_1)\zeta]s_1^+ \nonumber \\
\fl
&
+ (\gamma^\mathrm{in}_1 - \gamma^\mathrm{in}_1 \zeta) I
+ [ - (\gamma^\mathrm{in}_1 - \gamma^\mathrm{out}_1) + 2 \gamma^\mathrm{in}_1 \zeta ] n_1.
\label{eq_dual_H_1}
\end{eqnarray}

\subsubsection{Discussion for site $L$}

Here, we seek $\widetilde{H}^L$, which satisfies
\begin{eqnarray}
\exp_{q^{-2}} \left( q^{-\sum_{k=1}^L n_k} S^+ \right) H^L
= \widetilde{H}^L \exp_{q^{-2}} \left( q^{-\sum_{k=1}^L n_k} S^+ \right).
\end{eqnarray}
Different from site $1$,
all factors in the r.h.s.~in \eref{eq_factorization}
must be taken into the considerations.

Firstly, it is easy to confirm the following relation,
by using the similar discussion for site $1$ case;
\begin{eqnarray}
\exp_{q^{-2}}(s_L^+) H^L
= \left(-\gamma^\mathrm{in}_L s_L^- + (\gamma^\mathrm{in}_L + \gamma^\mathrm{out}_L) n_L \right) 
\exp_{q^{-2}}(s_L^+).
\end{eqnarray}
Secondly,
\begin{eqnarray}
\fl
\exp_{q^{-2}}(s_{L-1}^+ q^{-2n_L})
\left( - \gamma^\mathrm{in}_L s_L^- 
+ (\gamma^\mathrm{in}_L + \gamma^\mathrm{out}_L) n_L \right) \nonumber \\
\fl
= \left[
-\gamma^\mathrm{in}_L s_L^- + (\gamma^\mathrm{in}_L+\gamma^\mathrm{out}_L)n_L
- \gamma^\mathrm{in}_L(q^{-2}-1) s_L^- s_{L-1}^+
\right]
\exp_{q^{-2}}(s_{L-1}^+ q^{-2n_L}).
\end{eqnarray}
Although the successive calculations may become very complicated,
using the following fact that
\begin{eqnarray}
\fl
[s_i^+ q^{-2 \sum_{j=i+1}^L n_j}, s_L^- s_{k}^+]
&= s_i^+ q^{-2 \sum_{j=i+1}^L n_j} s_L^- s_{k}^+
- s_L^- s_{k}^+ q^{-2 \sum_{j=i+1}^L n_j} \nonumber \\
&= s_i^+ q^{-2 \sum_{j=i+1, j \neq k}^{L-1} n_j} q^{-2} s_L^- s_{k}^+
- s_L^- q^{-2} s_{k}^+ q^{-2 \sum_{j=i+1, j \neq k}^{L-1} n_j} \nonumber \\
&= 0,
\end{eqnarray}
we finally have the following result from the interchange: 
\begin{eqnarray}
\fl
\exp_{q^{-2}} \left( q^{-\sum_{k=1}^L n_k} S^+ \right) H^L \nonumber \\
\fl
= \left(
-\gamma^\mathrm{in}_L s_L^- + (\gamma^\mathrm{in}_L + \gamma^\mathrm{out}_L) n_L
- \sum_{i=1}^{L-1} \gamma^\mathrm{in}_L (q^{-2}-1) s_{L-i}^+ q^{-2\sum_{j=L-i+1}^{L-1}n_j} s_L^-
\right) 
\nonumber \\
\fl \quad
\times
\exp_{q^{-2}} \left( q^{-\sum_{k=1}^L n_k} S^+ \right),
\end{eqnarray}
and hence
\begin{eqnarray}
\fl
\widetilde{H}^L = 
-\gamma^\mathrm{in}_L s_L^- + (\gamma^\mathrm{in}_L + \gamma^\mathrm{out}_L) n_L
- \sum_{i=1}^{L-1} \gamma^\mathrm{in}_L (q^{-2}-1) s_{L-i}^+ q^{-2\sum_{j=L-i+1}^{L-1}n_j} s_L^-.
\label{eq_dual_H_L}
\end{eqnarray}

\subsection{Some comments for the derived results}

Note that the obtained quantum Hamiltonian
cannot be directly interpreted as the transition matrix
for the dual process;
the obtained quantum Hamiltonian does not satisfy the probability conservation law.
As in section~\ref{sec_SSEP_2},
it could be possible to use the Doi-Peliti formalism
to derive an adequate time-evolution operator for 
the dual stochastic process.
However, even from the dual quantum Hamiltonian
in \eref{eq_dual_H_1} and \eref{eq_dual_H_L},
the following facts are immediately obtained:
\begin{itemize}
\item The transition rates for the in-flow and out-flow 
of the particles on site $1$
will have the factor $\zeta = q^{-2 \sum_{j=2}^L n_j}$,
and hence the transition rates depend on the configurations.
It would be difficult to solve analytically the ASEP with such complicated boundary conditions.
\item Because the third term of $\widetilde{H}^L$ in \eref{eq_dual_H_L}
includes $s^+_{L-i} s^-_L$,
the long-range particle hopping
from site $L$ to site $L-i$ will occur in the dual process.
\end{itemize}

Hence, the dual process will become very complicated,
and then, at this stage,
there might be no benefit to consider the duality relation
for the ASEP with open boundary conditions.
Note that this fact can be revealed
by using the systematic derivation
introduced in the present paper.

\section{Concluding remarks}
\label{sec_discussions}

Using the systematic way based on the combination of the quantum spin language and the Doi-Peliti formalism,
the open boundary effects on the duality relations in the SSEP
and ASEP were discussed.
The systematic discussions give us
a general result for the SSEP with open boundaries;
it was clarified that not only the absorbing sites,
but also the copying sites are necessary in general.
As for the ASEP, it was clarified that the open boundary conditions
give complicated dual process, which would be difficult to solve analytically.
Hence, at this stage, there might be no merit to consider the duality relations
for the study of the ASEP with open boundary conditions.
However, as we saw in the present paper, 
heuristic ways would have little hope to find the complicated dual process;
the discussion in the present paper reveals the characteristics of the dual process
in a systematic way.

Up to now, the usage of the bosonic operators, i.e., the Doi-Peliti formalism,
has been basically restricted to the duality studies between the stochastic differential equations
and the birth-death processes \cite{Ohkubo2010,Ohkubo2013a},
in the context of the duality relations;
in the present paper, it was clarified that the bosonic operators
are also useful in the combination with the quantum spin language.
This technique could be hopeful to discuss duality relations
for other types of stochastic processes.

In the present paper, we focused on the expectations
$\widetilde{Q}_x$,
which has been investigated in other duality works for the ASEP.
It might be possible to obtain useful and simple dual processes
when we consider other physical quantities;
this is beyond the scope of the current work.

\section*{Acknowledgments}
The author is extremely grateful to Tomohiro Sasamoto,
with for useful discussions.
This work was supported in part by 
MEXT KAKENHI (Grants no. 25870339 and 16K00323)
and by the JSPS Core-to-Core program
``Non-equilibrium dynamics of soft-matter and information.''

\appendix

\section{Duality in SSEP with specific parameters}

If $2-\gamma^\mathrm{in}_i-\gamma^\mathrm{out}_i = 0$ for some $i$,
we must change the discussion slightly. 
For simplicity, assume here that 
$2-\gamma^\mathrm{in}_i-\gamma^\mathrm{out}_i = 0$ is satisfied for both $i =1$ and $i=L$;
it is straightforward to deal with more general cases.

In this case, we must go back to the boundary terms
in the dual original dual process in \eref{eq_SSEP_boundary_1}
and \eref{eq_SSEP_boundary_L};
\begin{eqnarray}
\widetilde{H}^1 = -\gamma^\mathrm{in}_1 s^-_1 + 2 n_1
= - 2 \left( \frac{1}{2} \gamma^\mathrm{in}_1 \right) s^-_1 + 2 n_1, \\
\widetilde{H}^L = -\gamma^\mathrm{in}_L s^-_L + 2 n_L 
= - 2 \left( \frac{1}{2} \gamma^\mathrm{in}_L \right) s^-_L + 2 n_L.
\end{eqnarray}
Hence, it is necessary to introduce only one sink site
at each boundary site;
we replace $\frac{1}{2}\gamma^\mathrm{in}_i$ with the annihilation operator
$a_i$ for the boundary site $i$.
In addition, the corresponding coherent state parameter
should be set to $\frac{1}{2}\gamma^\mathrm{in}_i$.
This boundary operator corresponds to the particle hopping
from the boundary site $i$ to the sink site attached to site $i$
``with rate $2$.''
Hence, the duality function becomes as follows:
\begin{eqnarray}
D\left(\eta,(\eta',\xi'_1, \xi'_L)\right) = 
&\left( \prod_{i \in \mathcal{S}; \, \eta'_i = 1} \eta_i \right)
\left( \frac{1}{2} \gamma^\mathrm{in}_1 \right)^{\xi'_1} 
\left( \frac{1}{2} \gamma^\mathrm{in}_L \right)^{\xi'_L} 
\end{eqnarray}
where $\xi'_i$ is the number of particles in sink site $i \in \{1,L\}$
in the dual stochastic process.

\section{Duality based on $\exp(q^{S^z} S+)$ in the ASEP with reflective boundaries}

From \eref{eq_wo_open_duality_initial},
we have
\begin{eqnarray}
\fl\frac{1}{(N-m)!} 
\sum_{1 \le y_1 < \dots < y_N \le L }
P_{\bm{y}}(t)
\langle x'_1, \dots, x'_m | (q^{S^z} S^+)^{N-m}
| y_1, \dots, y_N \rangle \nonumber \\
\fl
=
\frac{1}{(N-m)!} 
q^{-2\sum_{i=1}^m x'_i}
\sum_{1 \le x_1 < \dots < x_m \le L }
P'_{\bm{x}}(t)
q^{2\sum_{i=1}^m x_m}
\langle x_1, \dots, x_m | 
(q^{S^z} S^+)^{N-m} | y'_1, \dots, y'_N \rangle.
\nonumber \\
\end{eqnarray}
Here, we use the following projection state introduced in 
\cite{Imamura2011}:
\begin{eqnarray}
\sum_{\eta: \sum_{k=1}^L n_k = N} \langle \eta |
= \langle N | = C_N \langle 0 | (q^{S^z} S^+)^N,
\end{eqnarray}
where
\begin{eqnarray}
C_N = (q^{-2})^{\frac{LN}{4}}
\frac{(1-q^{-2})^N}{(1-q^{-2})\cdots(1- (q^{-2}))^N}.
\end{eqnarray}
From (2.27) in \cite{Imamura2011},
\begin{eqnarray}
\langle N | \widetilde{Q}_{x'_1} \cdots \widetilde{Q}_{x'_m} =
C_{N,m} \langle x'_1, \dots, x'_m | (X^-)^{N-m},
\label{eq_bra_IS}
\end{eqnarray}
where
\begin{eqnarray}
C_{N,m} = \frac{q^{-\frac{1}{2}(N-m)L} (1-q^{-2})^{N-m}}{(1-q^{-2}) \cdots (1-(q^{-2})^{N-m})}.
\end{eqnarray}
Hence, we have
\begin{eqnarray}
\fl
\sum_{1 \le y_1 < \dots < y_N \le L }
P_{\bm{y}}(t)
C_{N,m} \langle x'_1, \dots, x'_m | (q^{S^z} S^+)^{N-m}
| y_1, \dots, y_N \rangle \nonumber \\
\fl
=
q^{-2\sum_{i=1}^m x'_i}
\sum_{1 \le x_1 < \dots < x_m \le L }
P'_{\bm{x}}(t)
q^{2\sum_{i=1}^m x_m}
C_{N,m} \langle x_1, \dots, x_m | 
(q^{S^z} S^+)^{N-m} | y'_1, \dots, y'_N \rangle,
\end{eqnarray}
and finally
\begin{eqnarray}
\fl
\sum_{1 \le y_1 < \dots < y_N \le L }
P_{\bm{y}}(t)
\langle N | \widetilde{Q}_{x'_1} \cdots \widetilde{Q}_{x'_m}
| y_1, \dots, y_N \rangle \nonumber \\
\fl
=
q^{-2\sum_{i=1}^m x'_i}
\sum_{1 \le x_1 < \dots < x_m \le L }
P'_{\bm{x}}(t)
q^{2\sum_{i=1}^m x_m}
\langle N | \widetilde{Q}_{x_1} \cdots \widetilde{Q}_{x_m}
| y'_1, \dots, y'_N \rangle.
\end{eqnarray}
Note that the above derivation is based on the fact that
the number of particles is conserved.

\section{Verification of \eref{eq_w_open_quantity}}

\begin{eqnarray}
\fl
\sum_{N=0}^\infty
\langle x'_1, \dots, x'_m |
\rme_{q^{-2}} \left( (1-q^{-2}) q^{-\sum_{k=1}^L n_k} S^+ \right) 
\sum_{1 \le y_1 < \dots < y_N \le L} \phi_{\bm{y}}(t)
| y_1 , \dots, y_N \rangle 
\nonumber \\
\fl 
=
\sum_{N=0}^\infty
\langle x'_1, \dots, x'_m |
\sum_{n \ge 0} \frac{1}{ (q^{-2};q^{-2})_n }
\left( (1-q^{-2}) q^{-\sum_{k=1}^L n_k} S^+ \right)^n 
\sum_{1 \le y_1 < \dots < y_N \le L} \phi_{\bm{y}}(t)
| y_1 , \dots, y_N \rangle \nonumber \\
\fl
=\sum_{N=m}^\infty 
\frac{q^{- L(N-m)/2}}{  (q^{-2};q^{-2})_{N-m} } 
\langle x'_1, \dots, x'_m |
\left( (1-q^{-2}) q^{\sum_{k=1}^L s_k^z} S^+ \right)^{N-m} 
\sum_{1 \le y_1 < \dots < y_N \le L} \phi_{\bm{y}}(t)
| y_1 , \dots, y_N \rangle \nonumber \\
\fl
=\sum_{N=m}^\infty 
q^{- L(N-m)/2}
\frac{(1-q^{-2})^{N-m}}{  (q^{-2};q^{-2})_{N-m} } 
\frac{1}{C_{N,m}}
\langle N |
\widetilde{Q}_{x'_1} \cdots \widetilde{Q}_{x'_m}
\sum_{1 \le y_1 < \dots < y_N \le L} \phi_{\bm{y}}(t)
| y_1 , \dots, y_N \rangle.
\end{eqnarray}
Using the following equality,
\begin{eqnarray}
\fl
q^{- L(N-m)/2}
\frac{(1-q^{-2})^{N-m}}{  (q^{-2};q^{-2})_{N-m} } 
 \frac{1}{C_{N,m}} \nonumber \\
\fl
= q^{- L(N-m)/2} 
\frac{(1-q^{-2})^{N-m}}{(1-q^{-2}) (1-(q^{-2})^2) \cdots (1-(q^{-2})^{N-m})}
\frac{(1-q^{-2})\cdots (1-(q^{-2})^{N-m})}{q^{-(N-m)L/2} (1-q^{-2})^{N-m}}
\nonumber \\
\fl
= 1,
\end{eqnarray}
we have the usual expectation in \eref{eq_w_open_quantity}.

\vskip 5mm

\end{document}